\documentclass[10pt, preprint2]{emulateapj}
\usepackage{verbatim}
\usepackage[usenames, dvipsnames]{color}
\usepackage[pagebackref]{hyperref}

\newcommand{\thisstar}{HR 858}

\newcommand{\rsun}{\ensuremath{R_\sun}}
\newcommand{\msun}{\ensuremath{M_\sun}}
\newcommand{\lsun}{\ensuremath{L_\sun}}

\newcommand{\re}{\ensuremath{R_\earth}}
\newcommand{\fluxcgs}{10$^9$ erg s$^{-1}$ cm$^{-2}$}
\newcommand{\bjdtdb}{\ensuremath{\rm {BJD_{TDB}}}}
\newcommand{\fave}{\langle F \rangle}

\newcommand{\mearth}{\ensuremath{M_\earth}}

\newcommand{\kms}{\ensuremath{\rm km\,s^{-1}}}
\newcommand{\ms}{\ensuremath{\rm m\,s^{-1}}}

\newcommand{\Kepler}{\emph{Kepler}}
\newcommand{\TESS}{\emph{TESS}}
\newcommand{\Gaia}{\emph{Gaia}}
\newcommand{\tess}{\emph{TESS}}

\newcommand{\ut}{1}
\newcommand{\mitk}{2}
\newcommand{\usq}{8}
\newcommand{\cfa}{3}
\newcommand{\ames}{7}
\newcommand{\lco}{11}
\newcommand{\seti}{12}
\newcommand{\nexsci}{13}
\newcommand{\ucr}{14}
\newcommand{\louisville}{15}
\newcommand{\unc}{16}
\newcommand{\florida}{17}
\newcommand{\flatiron}{18}
\newcommand{\gmu}{19}
\newcommand{\ucsb}{20}
\newcommand{\open}{21}
\newcommand{\van}{22}
\newcommand{\fisk}{23}
\newcommand{\unsw}{24}
\newcommand{\nanj}{25}
\newcommand{\dunlap}{26}

\usepackage{xcolor}

\definecolor{my_color}{HTML}{3a18b1}

\definecolor{new_color}{HTML}{CF0000}
\definecolor{new_black}{HTML}{000000}

\newcommand*{\redit}{\textcolor{new_black}}
\newcommand*{\redittwo}{\textcolor{new_black}}

\shorttitle{Transiting super-Earths around HR 858}
\shortauthors{Vanderburg et al.}

\begin{document}

\title{\TESS\ Spots a Compact System of Super-Earths around the Naked-Eye Star HR 858}
\author{
Andrew Vanderburg\altaffilmark{\ut,27},   
Chelsea X. Huang\altaffilmark{\mitk,28},   
Joseph E. Rodriguez \altaffilmark{\cfa,29},  
Juliette C. Becker\altaffilmark{4,30,31}, 
George R.\ Ricker\altaffilmark{\mitk},   
Roland K.\ Vanderspek\altaffilmark{\mitk},   
David W.\ Latham\altaffilmark{\cfa},   
Sara Seager\altaffilmark{\mitk,5}, 
Joshua N.\ Winn\altaffilmark{6}, 
Jon M.\ Jenkins\altaffilmark{\ames}, 
Brett Addison\altaffilmark{\usq}, 
Allyson Bieryla\altaffilmark{\cfa},   
Cesar Brice\~no\altaffilmark{9}, 
Brendan P. Bowler\altaffilmark{\ut},   
Timothy M. Brown\altaffilmark{10, \lco}, 
Christopher J. Burke\altaffilmark{\mitk},   
Jennifer A. Burt\altaffilmark{\mitk,28},   
Douglas A.\ Caldwell\altaffilmark{\ames,\seti}, 
Jake T. Clark\altaffilmark{\usq},   
Ian Crossfield\altaffilmark{\mitk},   
Jason A.\ Dittmann\altaffilmark{5,32},   
Scott Dynes\altaffilmark{\mitk},   
Benjamin J. Fulton\altaffilmark{\nexsci}, 
Natalia Guerrero\altaffilmark{\mitk},   
Daniel Harbeck\altaffilmark{\lco}, 
Jonathan Horner\altaffilmark{\usq},   
Stephen R.\ Kane\altaffilmark{\ucr}, 
John Kielkopf\altaffilmark{\louisville}, 
Adam L. Kraus\altaffilmark{\ut},   
Laura Kreidberg\altaffilmark{\cfa, 33}, 
Nicolas Law\altaffilmark{\unc}, 
Andrew W. Mann\altaffilmark{\unc},   
Matthew W. Mengel\altaffilmark{\usq},   
Timothy D. Morton\altaffilmark{\florida,\flatiron}, 
Jack Okumura\altaffilmark{\usq},   
Logan A. Pearce\altaffilmark{\ut},  
Peter Plavchan\altaffilmark{\gmu}, 
Samuel N. Quinn\altaffilmark{\cfa}, 
Markus Rabus\altaffilmark{\lco, \ucsb}, 
Mark E. Rose\altaffilmark{\ames}, 
Pam Rowden\altaffilmark{\open}, 
Avi Shporer\altaffilmark{\mitk}, 
Robert J. Siverd\altaffilmark{\van}, 
Jeffrey C.\ Smith\altaffilmark{\ames, \seti},   
Keivan Stassun\altaffilmark{\van,\fisk}, 
C.G. Tinney\altaffilmark{\unsw},    
Rob Wittenmyer\altaffilmark{\usq},   
Duncan J. Wright\altaffilmark{\usq},   
Hui Zhang\altaffilmark{\nanj},   
George Zhou\altaffilmark{\cfa, 34},
Carl A. Ziegler\altaffilmark{\dunlap} 
}

\altaffiltext{\ut}{Department of Astronomy, The University of Texas at Austin, Austin, TX 78712, USA}
\altaffiltext{\mitk}{Department of Physics, and Kavli Institute for Astrophysics and Space Research, Massachusetts Institute of Technology, Cambridge, MA 02139, USA}
\altaffiltext{\cfa}{Center for Astrophysics $|$ Harvard and Smithsonian, Cambridge, MA 02138, USA}
\altaffiltext{4}{Department of Astronomy, University of Michigan, Ann Arbor, MI 48109, USA}
\altaffiltext{5}{Department of Earth and Planetary Sciences, MIT, 77 Massachusetts Avenue, Cambridge, MA 02139, USA}
\altaffiltext{6}{Department of Astrophysical Sciences, Princeton University, 4 Ivy Lane, Princeton, NJ 08544, USA}
\altaffiltext{\ames}{NASA Ames Research Center, Moffett Field, CA, 94035}
\altaffiltext{\usq}{Centre for Astrophysics, University of Southern Queensland, Toowoomba, QLD 4350, Australia}
\altaffiltext{9}{Cerro Tololo Inter-American Observatory, Casilla 603, La Serena, Chile}
\altaffiltext{10}{University of Colorado/CASA, Boulder, CO 80309, USA}
\altaffiltext{\lco}{Las Cumbres Observatory Global Telescope Network, Santa Barbara, CA 93117, USA}
\altaffiltext{\seti}{SETI Institute, Mountain View, CA 94043, USA}
\altaffiltext{\nexsci}{Caltech/IPAC-NExScI, 1200 East California Boulevard, Pasadena, CA 91125, USA}
\altaffiltext{\ucr}{Department of Earth and Planetary Sciences, University of California, Riverside, CA 92521, USA}
\altaffiltext{\louisville}{Department of Physics and Astronomy, University of Louisville, Louisville, KY 40292, USA}
\altaffiltext{\unc}{Department of Physics and Astronomy, University of North Carolina at Chapel Hill, Chapel Hill, NC 27599, USA}
\altaffiltext{\florida}{Department of Astronomy, University of Florida, Gainesville, FL, 32611, USA}
\altaffiltext{\flatiron}{Center for Computational Astrophysics, Flatiron Institute, 162 5th Ave., New York, NY 10010}
\altaffiltext{\gmu}{George Mason University, Fairfax, VA 22030, USA}
\altaffiltext{\ucsb}{Department of Physics, University of California, Santa Barbara, CA 93106-9530, USA}
\altaffiltext{\open}{School of Physical Sciences, The Open University, Milton Keynes MK7 6AA, UK}
\begin{abstract}
\textit{Transiting Exoplanet Survey Satellite} (\TESS) observations have revealed a compact multi-planet system around the sixth-magnitude star HR 858 (TIC 178155732, TOI 396), located 32 parsecs away. Three planets, each about twice the size of Earth, transit this slightly-evolved, late F-type star, which is also a member of a visual binary. Two of the planets may be in mean motion resonance. We analyze the \TESS\ observations, using novel methods to model and remove instrumental systematic errors, and combine these data with follow-up observations taken from a suite of ground-based telescopes to characterize the planetary system. The HR 858 planets are enticing targets for precise radial velocity observations, secondary eclipse spectroscopy, and measurements of the Rossiter-McLaughlin effect. 
\end{abstract}

\section{Introduction}
\label{sec:intro}
\footnotetext[\van]{Department of Physics and Astronomy, Vanderbilt University, Nashville, TN 37235 USA}
\footnotetext[\fisk]{Department of Physics, Fisk University, Nashville, TN 37208, USA}
\footnotetext[\unsw]{Exoplanetary Science at UNSW, School of Physics, UNSW Sydney, NSW 2052, Australia}
\footnotetext[\nanj]{School of Astronomy and Space Science, Key Laboratory of Modern Astronomy and Astrophysics in Ministry of Education, Nanjing University, Nanjing 210046, Jiangsu, China}
\footnotetext[\dunlap]{Dunlap Institute for Astronomy and Astrophysics, University of Toronto, Ontario M5S 3H4, Canada}
\footnotetext[27]{NASA Sagan Fellow}
\footnotetext[28]{Juan Carlos Torres Fellow}
\footnotetext[29]{Future Faculty Leaders Fellow}
\footnotetext[30]{NSF Graduate Research Fellow}
\footnotetext[31]{Leinweber Center for Theoretical Physics Graduate Fellow}
\footnotetext[32]{51 Pegasi b Postdoctoral Fellow}
\footnotetext[33]{Harvard Junior Fellow}
\footnotetext[34]{NASA Hubble Fellow}

\setcounter{footnote}{34}

The \Kepler\ space telescope was history's most prolific exoplanet-detecting tool \citep{borucki}. During its primary and extended K2 missions \citep{howell}, \Kepler\ searched over 500,000 stars across 5\% of the sky for small, periodic dimming events caused by transiting planets. \Kepler's survey revealed a stunning diversity of planets in terms of size \citep{fressin}, architecture \citep{lissauer}, and environment \citep{mann2017}, but due to the survey's design, most of \Kepler's discoveries orbit faint and otherwise anonymous stars hundreds or thousands of parsecs from Earth. Follow-up studies to characterize \Kepler's planets and investigate their detailed properties are limited by the host stars' faint apparent magnitudes. 


Now, the recently commissioned \textit{Transiting Exoplanet Survey Satellite} (\TESS, \citealt{ricker}) is beginning to identify analogs of the systems discovered by \Kepler, but around the nearest and brightest stars in the sky. Using four wide-angle cameras, \TESS\ is searching 80\% of the sky for transiting exoplanets during its two-year primary mission. Already, \TESS\ has discovered several new exoplanets around bright stars that are well-suited for follow-up observations \citep{huang2018, vanderspek}, and hundreds more \TESS\ planet candidates await confirmation.\footnote{\url{https://tess.mit.edu/alerts/}} 

\begin{figure*}[ht!]
\centering
	\includegraphics[width=6.5in]{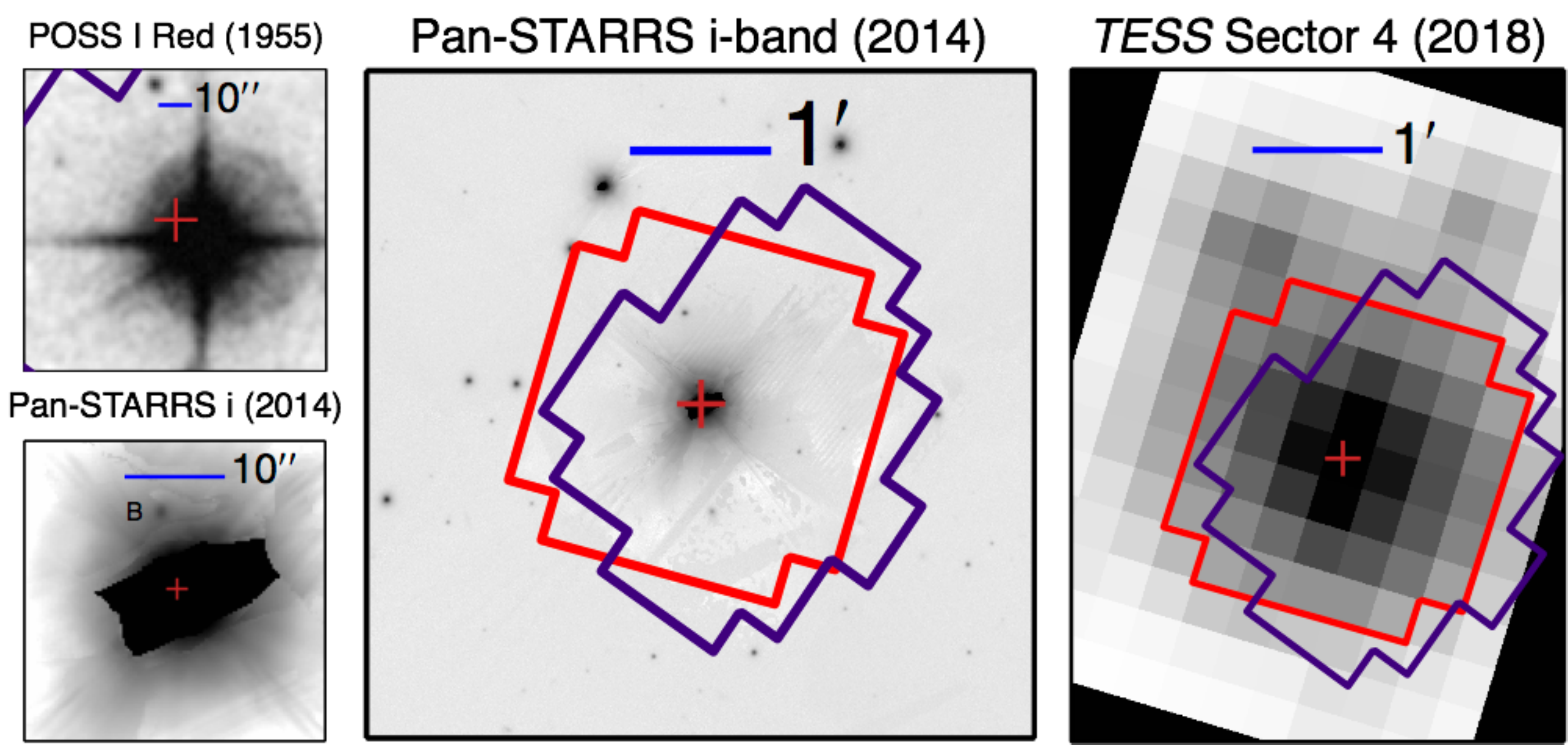}
    \caption{Images of the field surrounding \thisstar.
    \textit{Top left:}---From the
    first Palomar Observatory Sky Survey, obtained with
    a red-sensitive photographic emulsion in 1955. The red cross is the current position of \thisstar. 
    \textit{Bottom left:}---from the Pan-STARRS survey obtained with an i-band filter in 2014. The faint comoving companion \thisstar\ B is marked to the north of \thisstar. 
    \textit{Middle:}--- from the Pan-STARRS survey, over a wider field of view, obtained with an i-band filter in 2014.  Purple and red lines mark the boundary
    of the \TESS{} photometric apertures for Sectors 3 and 4, respectively.
    \textit{Right:}---Summed \TESS{} image. North is up and East is to the left in all the images.} 
    \label{fig:imaging}
\end{figure*}

Early in the mission, most \TESS\ planet discoveries were singly-transiting systems \citep{wang2019, rodriguez2019, nielsen2019}, \redit{but now that some stars have been observed for longer time baselines}, the survey is detecting transiting systems with increasingly complex architectures \citep{quinn2019, dragomir2019}. Here, we report the discovery of three super-Earths around the naked-eye star \thisstar. The planets are all about twice the size of Earth and have periods of 3.59, 5.98, and 11.23 days. \thisstar\ b and c orbit within 0.03\% of the 3:5 period ratio, and may be in true mean motion resonance. This compact and near-resonant architecture harkens back to the systems of tightly packed inner planets (STIPs) discovered by \Kepler, but \thisstar\ is hundreds to thousands of times brighter ($V$ = 6.4) than the hosts of those \Kepler\ systems. We describe our observations in Section \ref{sec:data}, our analysis to determine system parameters in Section \ref{sec:analysis}, and our efforts to show that the planet candidates are not false positives in Section \ref{sec:falsepositive}. We conclude by discussing the \thisstar\ system architecture and opportunities for follow-up observations in Section \ref{sec:discussion}. 



\section{Observations and Data Reduction}
\label{sec:data}

\subsection{TESS Photometry}\label{photometry}

\TESS\ observed \thisstar\ during the third and fourth sectors of its two-year-long primary mission, obtaining data from 20 September 2018 UT until 14 November 2018 UT. During Sector 4, \TESS\ saved and downlinked images of \thisstar\ every two minutes, standard procedure for the bright, nearby dwarf stars around which \TESS\ was specifically designed to discover planets. However, during Sector 3, \thisstar\ fell only a few pixels from the edge of the field of view, so (as for most of the sky) \TESS\ only downlinked co-added images with more-coarsely-sampled 30 minute cadence. 

Once the \TESS\ data were transmitted to Earth, we processed the data using two different sets of analysis tools in parallel: the MIT Quick Look Pipeline (QLP, \citealt{huang2019}) and the Science Processing Operations Center (SPOC, \citealt{Jenkins:2015, Jenkins:2016}) pipeline based at NASA Ames Research Center. After extracting light curves from the \TESS\ pixel data and searching for periodic signals, both pipelines identified the signatures of two transiting exoplanet candidates. These signals, which repeated every 3.59 and 5.98 days, were tested using standard diagnostics\footnote{These tests included searches for shallow secondary eclipses, differences in transit depth between even and odd-numbered transits, and shifts in \thisstar's apparent position during transit.} to determine whether the candidate transits were caused by some astrophysical or instrumental phenomenon other than a genuine system of transiting planets. We found no indication that these signals were false positives, and alerted the community to their existence via the MIT \TESS\ Alerts webpage.\footnote{\url{https://tess.mit.edu/alerts/}} We tentatively designated the planet candidates \thisstar\ b and c. 

\begin{figure*}[ht!]
\centering
	\includegraphics[width=6.5in]{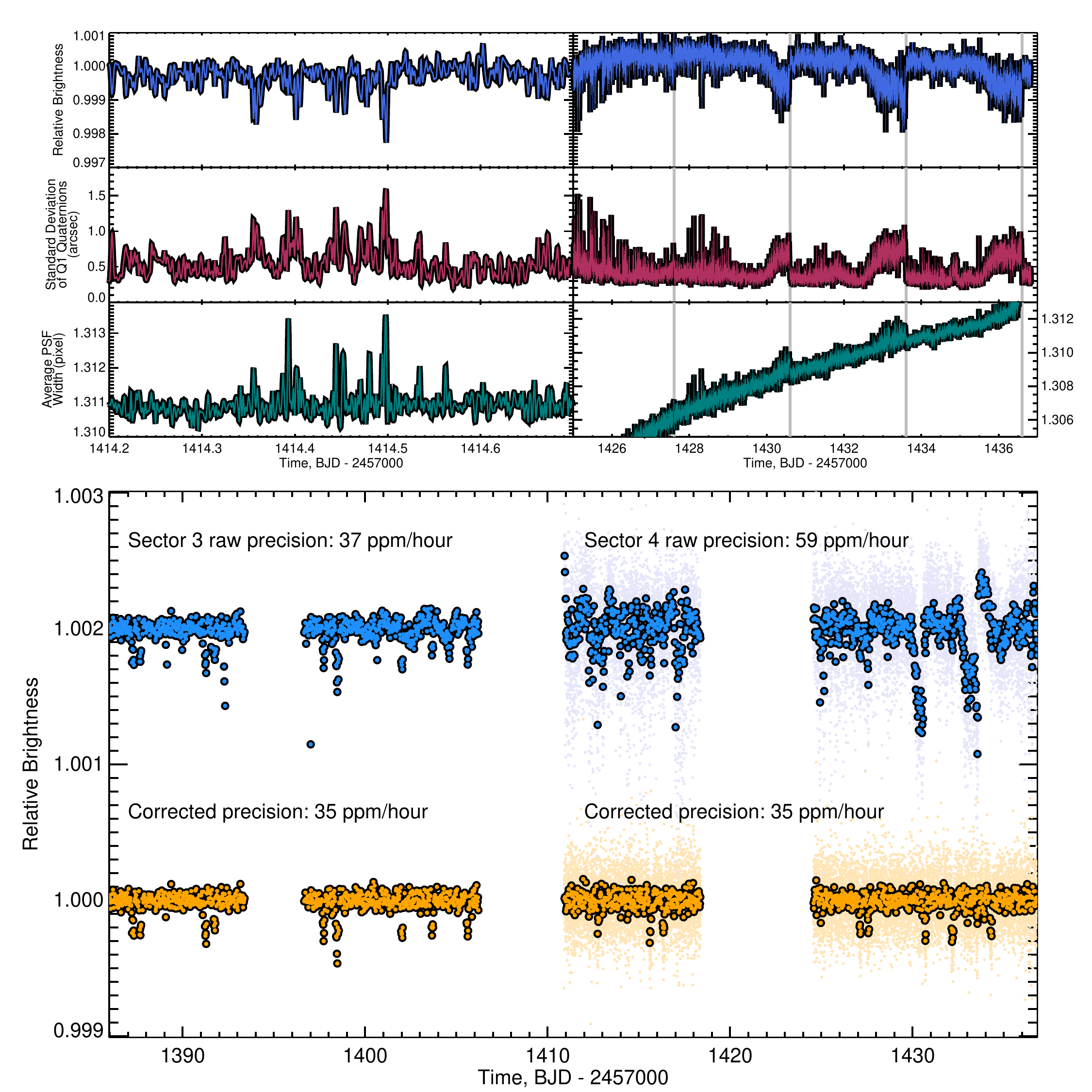}
    \caption{Systematic errors in the Sector 4 \tess\ light curve of \thisstar. Top block: Time series of measured brightness (top row), standard deviation of the Q1 quaternion component within each two-minute science exposure (middle row), and best-fit width of the \TESS\ PSF (bottom row). The time ranges shown in the two columns differ to emphasize the varying nature of the short and long timescale systematics. Times when the spacecraft underwent reaction wheel momentum dumps are shown as vertical grey lines. Note that the time series of PSF width shows a slow drift due to focus change after a spacecraft anomaly caused an onboard heater to activate. Bottom: \tess\ light curves of \thisstar\ before (blue) and after (orange) removal of systematic errors as described in Section \ref{photometry}. Faint points are individual two-minute exposures, and bold points are averages in 30 minute bins. The photometric precision is estimated by binning the light curve to one-hour exposures and calculating the point-to-point scatter.} 
    \label{fig:systematics}
\end{figure*}


Next, working from the calibrated pixel files,\footnote{We used a cutout from the online TESScut (\url{https://mast.stsci.edu/tesscut/}) tool for Sector 3 and the calibrated two-minute cadence target pixel files for Sector 4.} we re-extracted light curves from a series of both circular and irregularly shaped apertures \citep{vanderburg16}. We ultimately chose the apertures shown in Figure \ref{fig:imaging}, which minimized photometric scatter and contamination from two nearby stars. Systematic errors are present in the light curves from both sectors. Unlike \Kepler, whose instrumental systematics were dominated by changes in the spacecraft's focus \citep{jenkinslongcadence}, and K2, whose instrumental systematics were dominated by pointing drifts on timescales longer than single exposures \citep{vj14}, \TESS's instrumental systematics are dominated by pointing jitter on timescales shorter than an exposure. Figure \ref{fig:systematics} shows common features appearing in the \thisstar\ light curve, the width of the \TESS\ PSF (from a fit of the \tess\ images to a 2d Gaussian), and the intra-exposure scatter in engineering ``quaternion'' data.\footnote{The quaternion measurements are two-second-cadence vector time series that describe the spacecraft attitude based on observations of a set of guide stars. For each vector component (Q1,Q2,Q3), we take the standard deviation of all measurements within each two-minute science image. The quaternions are measured in each \TESS\ camera (in camera coordinates along the CCD row, column, and roll about the boresight), and are rotated into spacecraft coordinates (where the roll axis is pointing at the sky between Cameras 2 and 3). The quaternions are available online at \url{https://archive.stsci.edu/missions/tess/engineering/}.} These systematics are present on both short (exposure to exposure) and long ($\sim$day) timescales (which come from steady increases in the pointing scatter ahead of reaction wheel momentum dump events). 

We performed our own correction for the \TESS\ systematics. First, we ignored data where the SPOC quality flag was non-zero and during the following time intervals (where $t \equiv {\rm BJD} -2457000$):  $t<1385.96$ (while the \TESS\ operations team conducted tests on the spacecraft's attitude control system), $1393.4<t<1396.6361$,  $1406.25<t<1410.9054$, and $1423.5136<t<1424.5539$ (near \TESS's orbital perigee when Earthshine contaminated the aperture), and $1418.4915<t<1423.5136$ (when spacecraft/instrument communications were interrupted, shutting down the instrument, activating a heater, and introducing systematic trends).

We then treated the remaining short and long timescale systematic behavior separately. The long-timescale behavior gives rise to slow trends in the light curve with jumps each time the spacecraft resets the reaction wheel speeds by briefly firing its thrusters (a ``momentum dump''). We remove this behavior by fitting a basis spline (with robust outlier rejection and knots spaced roughly every 1.5 days) to the light curve and introducing discontinuities in the basis spline at the time of each momentum dump. Dividing this spline fit from the light curve effectively removes the long-timescale drifts. 

We treated the short-timescale behavior in the \TESS\ light curves differently between Sector 3 and Sector 4. During Sector 3, there are only a handful of exposures strongly affected by short-timescale pointing jitter. We simply exclude the 2\% of points with the widest measured PSF (indicating the largest intra-exposure pointing scatter), after removing slow drifts in the PSF width time series as done for the light curves (introducing discontinuities at momentum dumps). This cut corresponds roughly to excluding points with PSF width 7.5$\sigma$ larger than the high-pass-filtered median width, and removes all noticeable flux outliers from the light curve. \redit{This strategy is similar to that of \citet{fausnaugh2019}, who identified and removed anamolous points using the mean and standard deviation of the quaternion time series within exposures.}

The short-timescale systematic effects in the Sector 4 light curve were higher-amplitude and more pervasive, so instead of simply clipping strongly affected points from the time series, we opted to decorrelate the light curve against other time series. In Sector 4, instead of using the PSF width as a proxy for spacecraft motion, we worked with the less-noisy quaternion data,\footnote{We used the quaterions derived from Camera 2 (where \thisstar\ was observed) in camera coordinates. We converted the quaternion timestamps from spacecraft time to barycentric Julian date (BJD) towards \thisstar.} with long-term trends removed as done for the light curves and PSF width time series. We performed the decorrelation using a matrix-inversion least squares technique, iteratively removing 3$\sigma$ outliers from the fit until convergence. We experimented with decorrelating the light curve against different combinations of parameters including the averages and standard deviations of the (Q1, Q2, Q3) quaternions  within each exposure, averages and standard deviations of products of quaternions (Q1$\times$Q2, Q2$\times$Q3, Q1$\times$Q3), and various cotrending basis vectors used by the SPOC pipeline's Presearch Data Conditioning (PDC) module \citep{smith, stumpe}. We also experimented with decorrelating against higher (quadratic and cubic) orders of these time series. In the end, we found best results by decorrelating only against the standard deviation of the Q1 and Q2 quaternions and the seven cotrending vectors from PDC's band 3 (fast timescale) correction. The result of this decorrelation (and the long-timescale correction) on the Sector 4 \tess\ data are shown in the bottom panel of Figure \ref{fig:systematics}.

After producing light curves with systematic effects removed, we re-searched the light curve to look for additional transiting planet candidates. We searched the combined two-sector light curve (after binning the Sector 4 light curve to 30 minute cadence) with a Box-Least-Squares pipeline \citep{kovacs, vanderburg16}. In addition to recovering the two candidates identified by the QLP and SPOC pipelines, we detected a third convincing transit signal with a period of 11.23 days. \TESS\ detected three transits of this candidate: two in Sector 3, and one in Sector 4. After identifying the new candidate, we re-derived the systematics correction while excluding points taken during transits of all three planet candidates, and used this light curve in our analysis. 

We measured the centroid position of \thisstar\ in each \tess\ image and converted the measurements to time series in R.A. and DEC. The average changes in the position of \thisstar's centroid during each planet candidates' transits were consistent with zero (with precision of a few milliarcseconds). This confidently rules out the possibility that any star more than 40\arcsec\ away is the true source of the dimming events. 

\subsection{Archival and High Resolution Imaging} \label{sec:imaging}

We examined the region of sky around \thisstar\ using archival surveys and newly-obtained data (Figure \ref{fig:imaging}). Archival imaging from the Palomar Observatory Sky Survey (POSS) rules out background stars within about 6.5 magnitudes\footnote{Based on the lack of a visible bulge in \thisstar's saturated PSF and the size of the saturated PSFs of nearby 12$^{\rm th}$-13$^{\rm th}$ magnitude stars.} of \thisstar's brightness at its present day position, while images from the Pan-STARRS telescope identify seven stars besides \thisstar\ inside the TESS photometric apertures. All of these stars are at least 9 magnitudes fainter than \thisstar, and six of the seven are likely background objects. Parallax and proper motions observations from \Gaia\ DR2 \citep{gaiamission, gaiadr2} reveal that the nearest star to \thisstar\ (about 8\farcs4 to the northeast) is a co-moving companion (270 AU projected separation). \redittwo{The sky-projected velocity of \thisstar\ and the companion differ by only 2.108 $\pm$ 0.034 \kms, consistent with a bound orbit}, and analysis of the companion's spectral energy distribution \citep[SED, following][]{stassun2018} reveals it to be an M-dwarf with $T_{\rm eff} = 2800\pm300$ K and $R_\star = 0.17\pm0.04$ \rsun. The \Gaia\ observations of the comoving companion show large astrometric scatter; this may be due to systematic effects from the nearby, much brighter primary star, or it may be an indication that the comoving companion is itself an unresolved binary \citep{evans, rizzuto}. \redit{Some basic information about the comoving companion, which we call \thisstar\ B, is given in Table \ref{bigtable}.}

We also obtained a high-resolution I-band image of \thisstar\ with the HRCam speckle imager on the Southern Astrophysical Research (SOAR) telescope. The observations and analysis were conducted as described by \citet{Tokovinin2018}. Our observation was sensitive to nearly equal-brightness companions at separations of 0\farcs06 (1.8 AU projected distance) and fainter stars up to seven magnitudes fainter than \thisstar\ at larger (3\farcs15, 100 AU projected) separations. We detected no additional stars brighter than these contrast limits near \thisstar.  


\subsection{High Resolution Spectroscopy}\label{sec:spectroscopy}

We obtained high-resolution reconnaissance spectroscopy of \thisstar\ to determine spectroscopic parameters and rule out large radial velocity (RV) variations. \redit{We observed \thisstar\ twice with the Tillinghast Reflector Echelle Spectrograph (TRES\footnote{\url{www.sao.arizona.edu/html/FLWO/60/TRES/GABORthesis.pdf}}) on the 1.5m telescope at Fred L. Whipple Observatory, once with the CHIRON spectrograph on the 1.5m SMARTS telescope at Cerro Tololo Inter-American Observatory (CTIO), once with the echelle spectrograph on the 2.3m Australian National University (ANU) telescope at Siding \redittwo{Spring} Observatory, and seven times with the Network of Robotic Echelle Spectrographs (NRES, \citealt{nres0, nres, nres2}) operated by Las Cumbres Observatory \citep[LCO,][]{lco} from CTIO and South African Astronomical Observatory (SAAO).} The reconnaissance observations showed no large radial velocity variations or evidence for a composite spectrum. From the TRES data, we measured an absolute RV of 9.6 $\pm$ 0.1 \kms\ by cross-correlating the observed spectra with synthetic spectra derived from \citet{kurucz} atmosphere models and applying empirical corrections to shift the measured velocity to the IAU scale \citep{stefanik}. We found no evidence for large ($\sim$ \kms) RV variations that might indicate \thisstar\ is a close binary star. The measured absolute velocity is consistent with archival radial velocity measurements going back over a decade from the \Gaia\ mission \citep{gaiadr2}, Pulkuvo Observatory \citep{pulkuvo}, and the Geneva Copenhagen Survey \citep{casagrande2011}.

After our initial reconnaissance, we obtained 30 observations on 13 separate nights with the MINERVA-Australis telescope array at Mt. Kent Observatory in Queensland, Australia \citep{minervaaustralis} to place stronger limits on the transiting companions' masses. We measured radial velocities via least-squares analysis \citep{harpsterra} and corrected for spectrograph drifts with simultaneous Thorium Argon arc lamp observations. From these data, which showed scatter of about 14 \ms, we calculate upper limits (95\% confidence) on the masses of the three planet candidates around \thisstar\ of about 45 \mearth\ each using the RadVel package \citep{radvel}. \redittwo{Our radial velocity observations are summarized in Table \ref{rvtable}.}



We determined spectroscopic parameters from the TRES spectra using the Stellar Parameter Classification (SPC) code \citep{buchhave2012, buchhave2014} and found parameters ($T_{\rm eff} = 6201 \pm 50$, $\log{g_{\rm cgs}} = 4.19 \pm 0.10$, $\rm{[m/H]} =-0.14 \pm 0.08$)\footnote{[m/H] is the star's overall metallicity assuming HR 858's metals have the same relative proportions as in the Sun.} consistent with literature determinations \citep{gray2006, casagrande2011}. Our spectroscopic reconnaissance also found that \thisstar\ is rotating moderately rapidly. Following \citet{zhou2018}, we measured a projected rotational velocity of $v\sin i = 8.3 \pm 0.5$ \kms and a macroturbulent velocity of $v_{\rm mac} =  5.2 \pm 0.5$ \kms. \redit{An analysis of the NRES spectra using SpecMatch \citep{petigurathesis, petigura2017} yielded results ($T_{\rm eff} = 6199 \pm 100$, $\log{g_{\rm cgs}} = 4.3 \pm 0.1$, $\rm{[m/H]} =-0.10 \pm 0.07$) consistent with those from  TRES and SPC.}

\begin{deluxetable}{lccc}
\tablecaption{Summary of Radial Velocity Observations}
\tablewidth{0pt}
\tablehead{
  \colhead{Time} &   \colhead{RV }   &  \colhead{RV Error} &  \colhead{Instrument}    \\
  \colhead{BJD} &  \colhead{\kms}   &  \colhead{\kms} &  \colhead{}  }
\startdata
2458508.627 & 9.5450 & 0.1000 & TRES\\
2458510.649 & 9.5990 & 0.1000 & TRES\\
2458536.892 & 9.3000 & 0.5000 & ANU\\
2458532.543 & 8.1940 & 0.0140 & CHIRON\\
2458523.008 & 9.7820 & 0.0051 & MINERVA-Australis\\
2458523.016 & 9.7860 & 0.0049 & MINERVA-Australis\\
2458524.913 & 9.7892 & 0.0058 & MINERVA-Australis\\
2458524.927 & 9.7811 & 0.0058 & MINERVA-Australis\\
2458524.943 & 9.7773 & 0.0058 & MINERVA-Australis\\
2458524.957 & 9.7319 & 0.0056 & MINERVA-Australis\\
2458524.972 & 9.7726 & 0.0055 & MINERVA-Australis\\
2458526.974 & 9.8016 & 0.0058 & MINERVA-Australis\\
2458526.985 & 9.7963 & 0.0057 & MINERVA-Australis\\
2458528.948 & 9.7806 & 0.0046 & MINERVA-Australis\\
2458528.956 & 9.7688 & 0.0047 & MINERVA-Australis\\
2458528.964 & 9.7639 & 0.0047 & MINERVA-Australis\\
2458529.945 & 9.7676 & 0.0055 & MINERVA-Australis\\
2458529.952 & 9.7885 & 0.0056 & MINERVA-Australis\\
2458529.960 & 9.7886 & 0.0056 & MINERVA-Australis\\
2458530.943 & 9.7718 & 0.0054 & MINERVA-Australis\\
2458530.950 & 9.7779 & 0.0052 & MINERVA-Australis\\
2458531.943 & 9.7683 & 0.0051 & MINERVA-Australis\\
2458531.957 & 9.7761 & 0.0052 & MINERVA-Australis\\
2458533.947 & 9.7950 & 0.0059 & MINERVA-Australis\\
2458533.954 & 9.7917 & 0.0059 & MINERVA-Australis\\
2458535.982 & 9.8029 & 0.0057 & MINERVA-Australis\\
2458535.990 & 9.7817 & 0.0055 & MINERVA-Australis\\
2458536.940 & 9.7757 & 0.0056 & MINERVA-Australis\\
2458536.948 & 9.7740 & 0.0056 & MINERVA-Australis\\
2458537.954 & 9.7717 & 0.0054 & MINERVA-Australis\\
2458537.962 & 9.7793 & 0.0058 & MINERVA-Australis\\
2458538.955 & 9.7609 & 0.0048 & MINERVA-Australis\\
2458538.966 & 9.7682 & 0.0046 & MINERVA-Australis\\
2458509.048 & 9.8214 & 0.0096 & MINERVA-Australis\\
2458522.290 & 8.9689 & 0.7386 & LCO-SAAO\\
2458533.266 & 8.8251 & 0.2172 & LCO-SAAO\\
2458535.568 & 9.8806 & 0.1139 & LCO-CTIO\\
2458536.538 & 9.8718 & 0.1457 & LCO-CTIO\\
2458538.555 & 9.9197 & 0.1491 & LCO-CTIO\\
2458542.258 & 9.5613 & 0.2002 & LCO-SAAO\\
2458546.254 & 8.7488 & 0.1880 & LCO-SAAO\\
\enddata
\tablecomments{RVs from each instrument have not been corrected for instrumental offsets onto the same velocity system. Times have been converted to BJD\_TDB using routines written by \citet{eastmantimes}.}
\label{rvtable}
\end{deluxetable}

\begin{figure*}
\centering
	\includegraphics[width=6.5in]{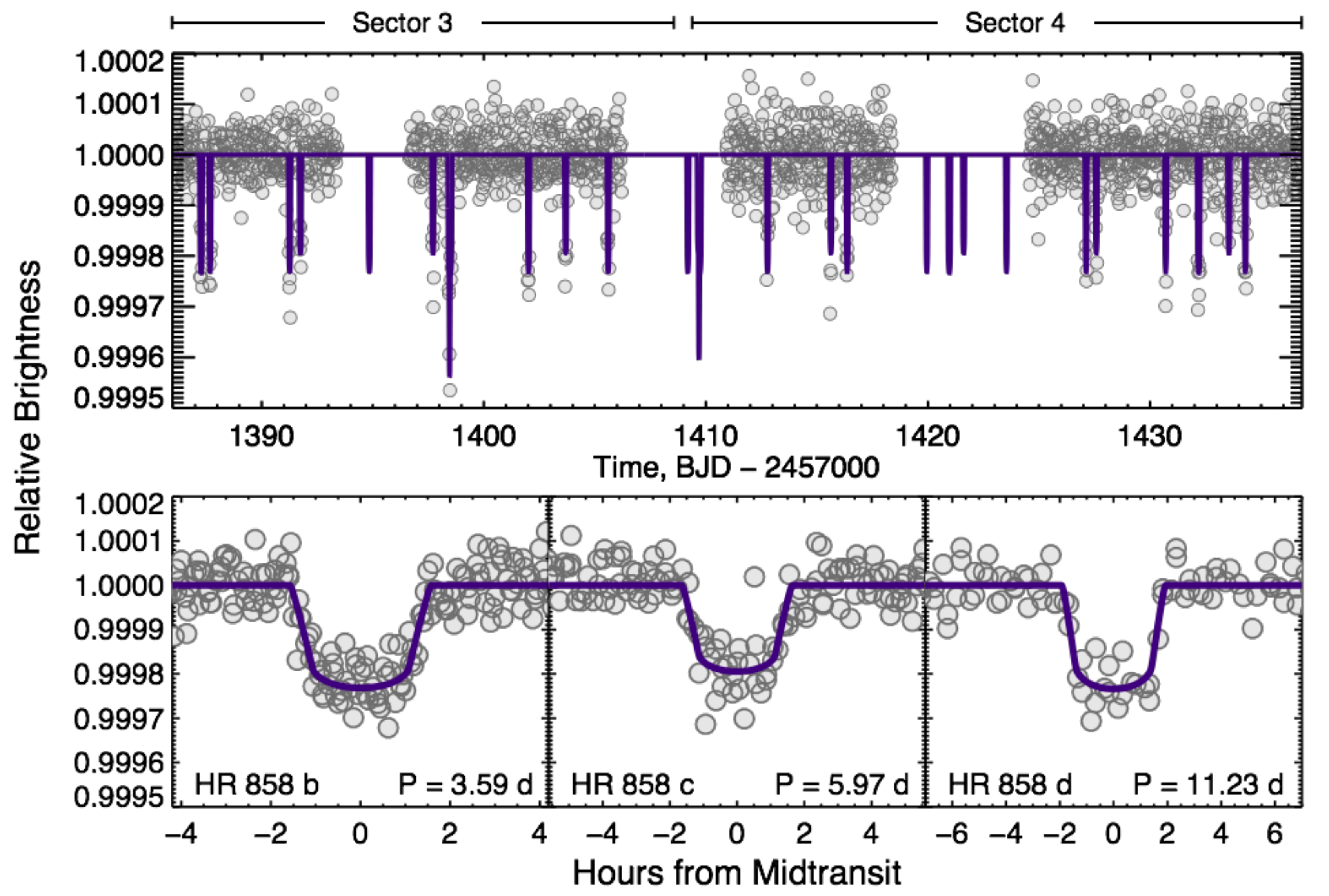}
    \caption{\TESS\ light curves of \thisstar. \textit{Top:} Full two-sector light curve. During Sector 3, \tess\ only downloaded images of \thisstar\ in its 30-minute cadence FFIs, but during Sector 4, \thisstar\ was pre-selected for observations at two-minute cadence. We bin the Sector 4 observations to an equivalent 30 minute cadence for visual clarity. \textit{Bottom:} Phase-folded \tess\ light curves of the three planets transiting \thisstar, with the Sector 4 observations again binned to 30 minute cadence for visual clarity. All analysis, including transit fitting, was performed on the unbinned two-minute-cadence Sector 4 light curve.} 
    \label{fig:lc}
\end{figure*}


\section{Determination of System Parameters}
\label{sec:analysis}

We determined system parameters using the EXOFASTv2 global modeling software \citep{eastman:2013, eastman:2017}. EXOFASTv2 uses Markov Chain Monte Carlo (MCMC) to explore a high-dimensional space in physical model parameters and determine best-fit values and uncertainties. We fit the two-sector \TESS\ light curve and a spectral energy distribution constructed from archival broadband photometry (listed in Table \ref{bigtable}). We imposed priors on spectroscopic parameters from TRES and the \Gaia\ parallax, and we enforced an upper limit on $V$-band extinction of $A_v < 0.04898$ mag from \citet{Schlegel:1998}. MIST isochrones \citep{mist} were used to constrain the stellar parameters. \redit{Each MCMC link's linear and quadratic limb darkening parameters were assigned by interpolating from \citet{claret2011} models at that link's surface gravity, effective temperature, and metallicity.} We ran the fit until convergence (defined as 1000 independent posterior draws after the chains all reached a Gelman-Rubin statistic less than 1.01). The results of our fit are given in Table \ref{bigtable}.

We cross-checked the EXOFASTv2 analysis with other less-comprehensive analyses in parallel. In particular, we fit for light curve and stellar parameters following \citet{huang2018}, and confirmed that our removal of low frequency variability and long-timescale systematics did not significantly affect the fit parameters. Another transit analysis that did not use constraints from the host star's parameters yielded the duration of transit ingress/egress, $t_{12}$ (or the time between the first and second transit contacts; see Figure 1 of \citealt{SeagerMallenOrnelas2003}) and the total transit duration, $t_{14}$ (from first to fourth contact). We also re-derived stellar parameters using an online interface\footnote{\url{http://stev.oapd.inaf.it/cgi-bin/param_1.3}} to fit the effective temperature, $V$-band magnitude and parallax with Padova models \citep{dasilva}, and using broadband photometry to fit the SED following \citet{stassun2018}; both analyses yielded results consistent with the EXOFASTv2 fit.

\section{False Positive Analysis}\label{sec:falsepositive}

While experience from the \Kepler\ mission has taught us that small planet candidates from space-based transit surveys are usually planets \citep{mortonjohnson}, especially those in multi-transiting systems \citep{lissauermultis}, careful analysis is required to rule out false positive scenarios. During the \Kepler\ and K2 eras, it became common to ``statistically validate'' planet candidates using tools like \texttt{vespa} \citep{morton12, morton15}, \texttt{BLENDER} \citep{blender}, and \texttt{PASTIS} \citep{pastis} which quantify the likelihood that the any given signal arises from a false positive.

Planet candidates discovered by \tess\ often have advantages over candidates from \Kepler/K2, which can make it possible to rule out some or all false positive scenarios categorically, rather than calculating probabilities based on false positive population models. In particular, most \tess\ planet candidates are observed at 2 minute cadence, so we can precisely measure ingress/egress times, and many \tess\ targets are nearby and have high proper motion, so it is possible to show that background stars cannot cause the transit signals. 

For \thisstar, we take advantage of both approaches. We consider the following false positive scenarios for one or more of the transit signals around \thisstar: 
\begin{enumerate}
    \item \textit{\thisstar\ is an eclipsing binary:} Our radial velocity observations from MINERVA-Australis and TRES rule out this scenario (Section \ref{sec:spectroscopy}).
    \item \textit{Light from an unassociated eclipsing binary or transiting planet system is blended with \thisstar:} If the transit signal comes from a star other than \thisstar, the observed transit depth $\delta$ will be:
    
    \begin{equation}
        \delta \simeq \left ( \frac{R_{p, {\rm true}}}{R_\star} \right )^2 \frac{F_{\rm source}}{F_{\rm total}}
    \end{equation}
    
    \noindent where ${R_{p, {\rm true}}}/{R_\star}$ is the true radius ratio of the transiting/eclipsing body on the source star, and ${F_{\rm source}}/{F_{\rm total}}$ is the fraction of the flux the source star contributes to the \tess\ light curve. 
     The ratio of the transit ingress/egress duration, $t_{12}$, to the duration from first to third contact ($t_{13}\equiv t_{14} - t_{12}$) constrains the radius ratio of the transit source regardless of any diluting flux \citep[from][Equation 21]{SeagerMallenOrnelas2003}: 
    
    \begin{equation}
        \frac{R_{p, {\rm true}}}{R_\star }\leq \frac{t_{12}}{t_{13}}
    \end{equation}
     
    \noindent We calculate the magnitude difference $\Delta m$ between \thisstar\ and the faintest companion which could cause the transit signals we see: 
    \begin{equation}
    \delta \lesssim \left ( \frac{t_{12}}{t_{13}} \right )^2 \frac{F_{\rm source}}{F_{\rm total}} \approx \left ( \frac{t_{12}}{t_{13}} \right )^2 10^{-0.4 \Delta m}
    \end{equation}
    \begin{equation}\label{deltam}
        \Delta m \lesssim 2.5 \log_{10}\left ( \frac{t_{12}^2}{t_{13}^2 \delta} \right )
    \end{equation}
    
    \noindent Using Equation \ref{deltam} and our transit analysis (Section \ref{sec:analysis}, we find $\Delta m \lesssim$ 4.5, 5.9, and 6.1 magnitudes for \thisstar\ b, c, and d respectively (95\% confidence). Analysis of the \TESS\ image centroids shows that the source of the transits must be within 40\arcsec\ of \thisstar\ (Section \ref{photometry}), and archival imaging (Section \ref{sec:imaging}) shows no stars both close enough and bright enough to contribute the transit signals, including at \thisstar's present-day position, ruling out background false positive scenarios. 
    
    \item \textit{Light from a physically associated companion which is an eclipsing binary or transiting planet system is blended with \thisstar:}  The comoving companion \thisstar\ B is too faint \redit{(\Gaia\ Rp = 14.5, $\Delta$Rp = 8.6 mag)} to contribute the transits based on our $\Delta m$ constraints, and there is no evidence of any brighter resolved companions in speckle/archival imaging or any unresolved companion causing an RV acceleration. False positive scenarios involving bound companions to \thisstar\ are therefore unlikely, but we can not conclusively rule out the possibility that \thisstar\ has an undetected companion bright enough to contribute the transits. 
\end{enumerate}

\begin{figure*}
\centering
	\includegraphics[width=6.5in]{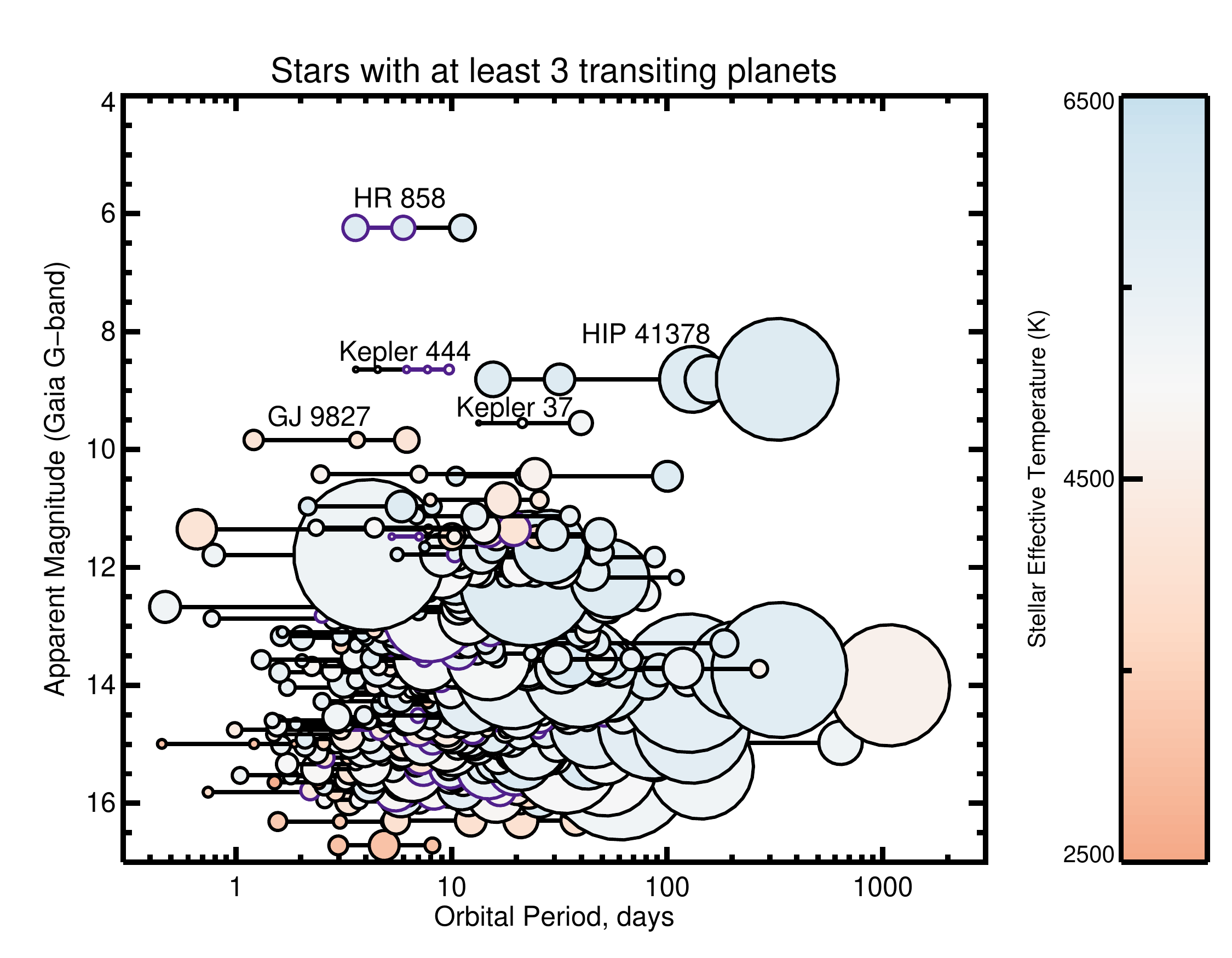}
    \caption{\thisstar\ in the context of other known transiting exoplanet systems. Shown plotted are all known systems with at least three transiting planets, as a function of the host stars' apparent brightness at visible wavelengths. The planets within each system are connected together with a horizontal line, and the planet radii and host star effective temperatures are encoded by the size and color of the symbols, respectively. We identified all pairs of planets within 1\% of first order (1:2, 2:3, 3:4, 4:5, 5:6, 6:7) mean motion resonances, and within 0.1\% of second-order (1:3, 3:5, 5:7) mean motion resonances and connected these planets together with purple lines. \thisstar\ stands out as the brightest of all known three-transiting-planet systems, while its compact near-resonant architecture is reminiscent of the population of compact multi-planet systems discovered by \Kepler.} 
    \label{fig:multisystems}
\end{figure*}

Since we cannot rule out all false positive scenarios involving physically associated companions to \thisstar, we use \texttt{vespa} to evaluate the probability of these false positive scenarios. Using the \tess\ light curve of each planet candidate and constraints from spectroscopy and imaging, \texttt{vespa} finds low false positive probabilities (FPP) for all three planet candidates (FPP$<10^{-3}$ for each candidate), so we consider \thisstar\ b, c, and d to be validated planets. 





\section{Discussion}
\label{sec:discussion}

\thisstar\ is one of the brightest stars known to host transiting planets, trailing only HD 219134 \citep{Motalebi:2015}, $\pi$ Mensae \citep{huang2018}, and 55 Cancri \citep{winn2011}. Transiting planets around stars this bright afford rich opportunities for detailed characterization, including mass measurements through precise RV observations, secondary eclipse spectroscopy with the \textit{James Webb Space Telescope}\footnote{\redit{The \texttt{PandExo} tool predicts that NIRCam observations (with a grism and the F444W filter) of a single secondary eclipse of \thisstar\ b should yield an 11$\sigma$ detection over the full bandpass and $\approx$ 30\% precision in 100 nm spectral bins. Despite \thisstar\ being near JWST's bright limits, the simulated observing efficiency was 67\% using the SUBGRISM64, frametime = 0.34 seconds read-out mode, and the star did not saturate the detectors.}}, and measurements of the alignment of the planetary orbits and stellar spin axis via the Rossiter-McLaughlin effect \citep{rossiter, mclaughlin} or Doppler Tomography.\footnote{\redit{Though \thisstar's moderate rotation complicates RV observations, it is possible to measure precise RVs of even more rapidly rotating stars \citep{barros}. Early RV observations of \thisstar\ }indicate it is possible to achieve precision of a few \ms\ (D. Gandolfi and D. Anderson, priv. comm.), similar to the expected 1-2 \ms\ amplitude of the RV orbits and Rossiter-McLaughlin signals.} \thisstar\ stands out even among the brightest known transiting systems because of its multiplicity; the next-brightest star known to host 3 transiting planets is 9 times fainter than \thisstar. (see Figure \ref{fig:multisystems}).





From \Kepler, we know of many examples of compact, multi-transiting, and coplanar systems, but relatively few of these systems are in true mean motion resonances \citep{fabrycky2014}. The \thisstar\ system could be one of the exceptions to this rule; \thisstar\ b and c may be in a true 3:5 mean motion resonance. We assessed these planets' resonant state by randomly drawing 50 sets of initial orbital parameters from the EXOFASTv2 posterior probability distributions and performing N-body integrations for $10^5$ years using the \texttt{Mercury6} \citep{mercury6} code. We used a hybrid symplectic and Bulirsch-Stoer integrator, with a time-step of 90 minutes and energy conservation kept to 1 part in $10^{8}$ or better. The simulations include the stellar quadrupole field due to rotation as a $J_{2}$ moment, which we estimated to be $J_{2} \approx 10^{-6}$ by modeling the star as a $n = 3$ polytrope \citep{Lanza2011, Batygin2013} with our derived mass/radius/rotational velocity. About one third of the simulated system realizations show at least some evidence of mean motion resonance: \redit{20\% of the realizations exhibited librating resonance angles for the entire duration of the simulation, while another 14\% exhibited librating resonance angles some of the time.}  Long-term RV and/or transit monitoring will help determine the resonant state of these planets; lower eccentricities for planets b and c\footnote{\redit{The 3:5 mean motion resonance, in particular, is both easier to generate during disk migration (Quillen 2006) and more easily maintained in the presence of nearby planetary perturbers when the system eccentricities are low. A more narrow libration width, characteristic at lower eccentricities, presents fewer opportunities for a nearby perturbing planet to disrupt the resonance.}}, and weaker perturbations from the outer planet d\footnote{In our simulations, we found evidence that lower masses and eccentricities for planet d increased the likelihood of planets b and c being in resonance.} should make a resonance more likely. 

The comoving stellar companion, \thisstar\ B, adds further intrigue to the system's architecture. The \Gaia\ proper motion measurements for the primary and secondary differ by 13.9 $\pm$ 0.2 mas\,yr$^{-1}$. If we interpret this discrepancy as relative orbital motion between the two stars (and not systematics due to the large brightness contrast or unresolved orbital motion if \thisstar\ B is indeed itself a close binary), the orbit of \thisstar\ B about \thisstar\ A must be misaligned from the orbits of the transiting system by at least 40 degrees.\footnote{As determined by a fit of the binary orbit using code available at \url{https://github.com/logan-pearce/LOFTI} (Pearce et al. \textit{in prep}).} If true, \thisstar\ B could have torqued \thisstar's planet-forming disk, causing a misalignment between the stellar spin axis and the transiting super-Earths' orbits. In particular, HR 858 B's mass and projected separation appear to put the system in a regime where the timescale for stellar spin axis realignment would be longer than the disk dissipation timescale, potentially ``freezing in'' the misalignment \citep{batygin, spaldingbatygin}. Future monitoring of the \thisstar\ A/B binary orbit should confirm its misalignment with the transit system and determine whether these mechanisms could create a spin/orbit misaligned multi-planetary system (that could be identified via Rossiter-McLaughlin observations of \thisstar\ b, c, or d).






Though the \tess\ prime mission survey is only about 25\% complete, there may not be many new transiting planets around stars brighter than \thisstar\ left to discover. Pre-launch estimates of the \tess\ planet yield \citep{sullivanyield, barclayyield, huangyield} predicted a handful of planet discoveries around naked eye stars, and so far only \thisstar\ and $\pi$ Mensae have fit this description. \thisstar\ will thus likely retain its privileged position as one of the brightest transit hosts in the sky and most favorable systems for detailed study. 



\acknowledgments
We thank Jen Winters and Michael Fausnaugh for helpful discussions. We acknowledge the use of public \TESS\ Alert data from pipelines at the \TESS\ Science Office and at the \TESS\ Science Processing Operations Center.
Funding for the \TESS\ mission is provided by NASA's Science Mission directorate. Resources supporting this work were provided by the NASA High-End Computing (HEC) Program through the NASA Advanced Supercomputing (NAS) Division at Ames Research Center for the production of the SPOC data products.
AV's work was performed under contract with the California Institute of Technology / Jet Propulsion Laboratory funded by NASA through the Sagan Fellowship Program executed by the NASA Exoplanet Science Institute.
CXH and JB acknowledge support from MIT's Kavli Institute as Torres postdoctoral fellows.
JAD and JNW acknowledge support from the Heising--Simons Foundation.
JER is supported by the Harvard Future Faculty Leaders Postdoctoral fellowship. 
J.C.B is supported by the NSF Graduate Research Fellowship Grants No. DGE 1256260 and a graduate fellowship from the Leinweber Center for Theoretical Physics.
Some of the data presented in this paper were obtained from the Mikulski Archive for Space Telescopes (MAST). STScI is operated by the Association of Universities for Research in Astronomy, Inc., under NASA contract NAS5--26555. Support for MAST for non--HST data is provided by the NASA Office of Space Science via grant NNX13AC07G and by other grants and contracts. This work has made use of data from the European Space Agency (ESA) mission {\it Gaia} (\url{https://www.cosmos.esa.int/gaia}), processed by the {\it Gaia} Data Processing and Analysis Consortium (DPAC, \url{https://www.cosmos.esa.int/web/gaia/dpac/consortium}). Funding for the DPAC
has been provided by national institutions, in particular the institutions participating in the {\it Gaia} Multilateral Agreement.
MINERVA-Australis is supported by Australian Research Council LIEF Grant LE160100001, Discovery Grant DP180100972, Mount Cuba Astronomical Foundation, and institutional partners University of Southern Queensland, UNSW Australia, MIT, Nanjing University, George Mason University, University of Louisville, University of California Riverside, University of Florida, and University of Texas at Austin. This work makes use of observations from the LCOGT network.
This research has made use of NASA's Astrophysics Data System and the NASA Exoplanet Archive, which is operated by the California Institute of Technology, under contract with the National Aeronautics and Space Administration under the Exoplanet Exploration Program. The National Geographic Society--Palomar Observatory Sky Atlas (POSS-I) was made by the California Institute of Technology with grants from the National Geographic Society. The Oschin Schmidt Telescope is operated by the California Institute of Technology and Palomar Observatory.


\textit{Facilities:} 
\facility{\TESS}, 
\facility{FLWO:1.5m (TRES)},
\facility{MINERVA-Australis}, \facility{SOAR (HRCAM)}, \facility{LCO:1m (NRES)}, \facility{ATT (echelle)}, \facility{CTIO:1.5m (CHIRON)}.

\redit{\textit{Software:} IDL Astronomy Library \citep{idlastronomylibrary}, EXOFASTv2 \citep{eastman:2013, eastman:2017}, Mercury6 \citep{mercury6}, vespa \citep{morton12, morton15}, Orbits for the Impatient (\citealt{blunt}, Pearce et al. \textit{in prep}), RadVel \citep{radvel}, numpy \citep{np}, pandas \citep{pandas}, matplotlib \citep{plt}, Pandexo \citep{pandexo}.}


\begin{deluxetable*}{lcccc}
\tablecaption{HR 858 System Parameters}
\tabletypesize{\tiny}
\tablehead{\colhead{~~~Parameter} & \colhead{Units} & \multicolumn{3}{c}{Values}}
\centering
\startdata
\multicolumn{2}{l}{Identifying Information}\\
\hline\\
\multicolumn{5}{l}{~~~~HR 858, HD 17926, HIP 13363, TIC 178155732, TOI 396}\\
\multicolumn{2}{l}{~~~~\Gaia\ DR2 Source ID 5064574720469473792}\\
\\
~~~~R.A\dotfill &Right Ascension (J2000)\dotfill &02:51:56.25 \\ 
~~~~DEC\dotfill &Declination (J2000)\dotfill &-30:48:52.3 \\ 
~~~~PM$_{\rm RA}$\dotfill &Proper Motion in Right Ascension (mas\,yr$^{-1}$)\dotfill &$123.229\pm0.070$ \\
~~~~PM$_{\rm DEC}$.\dotfill &Proper Motion in Declination (mas\,yr$^{-1}$)\dotfill &$105.788\pm0.151$ \\
~~~~$\varpi$\dotfill &Parallax (mas)\dotfill &$31.256\pm0.070$\\
\\
~~~~$B_{\rm T}$\dotfill &Tycho B-band Magnitude\dotfill &$6.956\pm0.015$ \\
~~~~$V_{\rm T}$\dotfill &Tycho V-band Magnitude\dotfill &$6.438\pm0.010$ \\
~~~~$J$\dotfill &2MASS J-band Magnitude\dotfill &$5.473\pm0.030$ \\
~~~~$H$\dotfill &2MASS H-band Magnitude\dotfill &$5.225\pm0.030$ \\
~~~~$Ks$\dotfill &2MASS K-band Magnitude\dotfill &$5.149\pm0.020$ \\
~~~~$W1$\dotfill &WISE Band 1 Magnitude\dotfill &$5.098\pm0.232$ \\
~~~~$W2$\dotfill &WISE Band 2 Magnitude\dotfill &$4.941\pm0.094$ \\
~~~~$W3$\dotfill &WISE Band 3 Magnitude\dotfill &$5.171\pm0.014$ \\
~~~~$W4$\dotfill &WISE Band 4 Magnitude\dotfill &$5.100\pm0.029$ \\

\\\hline

\multicolumn{5}{l}{Identifying Information and Photometric Properties for Co-moving Companion HR 858 B}\\
\hline\\
\multicolumn{2}{l}{~~~~\Gaia\ DR2 Source ID 5064574724768583168}\\
\\
~~~~R.A\dotfill &Right Ascension (J2000)\dotfill & 02:51:56.41\\
~~~~DEC\dotfill &Declination (J2000)\dotfill & -30:48:44.2\\
~~~~PM$_{\rm RA}$\dotfill &Proper Motion in Right Ascension (mas\,yr$^{-1}$)\dotfill &$137.125\pm0.213$ \\
~~~~PM$_{\rm DEC}$.\dotfill &Proper Motion in Declination (mas\,yr$^{-1}$)\dotfill &$105.865\pm0.302$ \\
~~~~$\varpi$\dotfill &Parallax (mas)\dotfill &$32.301\pm0.167$\\
\\
~~~~$G$\dotfill &\Gaia\ G-band Magnitude\dotfill &$16.0464\pm0.05$ \\
~~~~$\rm Bp$\dotfill &\Gaia\ Bp-band Magnitude\dotfill &$17.0142\pm0.1$ \\
~~~~$\rm Rp$\dotfill &\Gaia\ Rp-band Magnitude\dotfill &$14.4812\pm0.1$ \\
~~~~$i$\dotfill &Pan-STARRS $i$-band Magnitude\dotfill &$14.4611\pm0.05$ \\
~~~~$z$\dotfill &Pan-STARRS $z$-band Magnitude\dotfill &$14.1671\pm0.05$ \\
~~~~$y$\dotfill &Pan-STARRS $y$-band Magnitude\dotfill &$13.0732\pm0.08$ \\

\\\hline
\multicolumn{2}{l}{Observed Stellar Parameters}\\
\hline\\

~~~~$\log{g}$\dotfill &Spectroscopic surface gravity (cgs)\dotfill &$4.19\pm0.1$ \\
~~~~$T_{\rm eff}$\dotfill &Effective Temperature (K)\dotfill & $6201\pm50$ \\
~~~~$[{\rm Fe/H}]$\dotfill &Metallicity (dex)\dotfill & $-0.14\pm 0.08$\\
~~~~$v\sin{i}$\dotfill &Projected rotational velocity (\kms)\dotfill &$8.3\pm0.5$\\


\\\hline
\multicolumn{2}{l}{Derived Stellar Parameters}\\
\hline\\
~~~~$M_\star$\dotfill &Mass (\msun)\dotfill &$1.145^{+0.074}_{-0.080}$\\
~~~~$R_\star$\dotfill &Radius (\rsun)\dotfill &$1.310^{+0.024}_{-0.022}$\\
~~~~$L_\star$\dotfill &Luminosity (\lsun)\dotfill &$2.348^{+0.069}_{-0.071}$\\
~~~~$\rho_\star$\dotfill &Density (cgs)\dotfill &$0.717^{+0.064}_{-0.063}$\\
~~~~$\log{g}$\dotfill &Model-derived surface gravity (cgs)\dotfill &$4.262^{+0.032}_{-0.036}$\\
~~~~$u_{1}$\dotfill &\TESS-band linear limb-darkening coeff \dotfill &$0.227\pm0.034$\\
~~~~$u_{2}$\dotfill &\TESS-band quadratic limb-darkening coeff \dotfill &$0.295\pm0.035$\\
\\\hline
\multicolumn{2}{l}{Planetary Parameters:}&b &c&d\\
\hline\\
~~~~$P$\dotfill &Period (days)\dotfill &$3.58599\pm0.00015$&$5.97293^{+0.00060}_{-0.00053}$&$11.2300^{+0.0011}_{-0.0010}$\\
~~~~$R_P$\dotfill &Radius (\re)\dotfill &$2.085^{+0.068}_{-0.064}$&$1.939\pm0.069$&$2.164^{+0.086}_{-0.083}$\\
~~~~$T_C$\dotfill &Time of conjunction (\bjdtdb)\dotfill &$2458409.18969^{+0.00084}_{-0.00083}$&$2458415.6344^{+0.0022}_{-0.0014}$&$2458409.7328^{+0.0020}_{-0.0018}$\\
~~~~$a$\dotfill &Semi-major axis (AU)\dotfill &$0.0480^{+0.0010}_{-0.0011}$&$0.0674^{+0.0014}_{-0.0016}$&$0.1027^{+0.0022}_{-0.0025}$\\
~~~~$i$\dotfill &Inclination (Degrees)\dotfill &$85.50^{+1.5}_{-0.50}$&$86.23\pm0.26$&$87.43^{+0.18}_{-0.19}$\\
~~~~$e$\dotfill &Eccentricity (95\% Confidence) \dotfill &$<0.30$&$<0.19$&$<0.28$\\
~~~~$T_{eq}$\dotfill &Equilibrium temperature (K)\dotfill &$1572^{+22}_{-19}$&$1326^{+18}_{-16}$&$1075^{+15}_{-13}$\\
~~~~$R_P/R_\star$\dotfill &Radius of planet in stellar radii \dotfill &$0.01460\pm0.00035$&$0.01358^{+0.00038}_{-0.00039}$&$0.01514^{+0.00050}_{-0.00049}$\\
~~~~$a/R_\star$\dotfill &Semi-major axis in stellar radii \dotfill &$7.87^{+0.23}_{-0.24}$&$11.06^{+0.32}_{-0.34}$&$16.85^{+0.49}_{-0.51}$\\
~~~~$d/R_\star$\dotfill &Planet/star separation at mid transit \dotfill &$7.29^{+0.83}_{-1.1}$&$10.88^{+0.64}_{-0.81}$&$15.9^{+1.8}_{-2.3}$\\
~~~~$\delta$\dotfill &Transit depth $\left (R_p/R_\star \right )^2$\dotfill &$0.000213\pm0.000010$&$0.000184\pm0.000011$&$0.000229\pm0.000015$\\
~~~~$T_{14}$\dotfill &Total transit duration (days)\dotfill &$0.1129\pm0.0016$&$0.1209^{+0.0045}_{-0.0030}$&$0.1431^{+0.0042}_{-0.0038}$\\
~~~~$b$\dotfill &Transit Impact parameter \dotfill &$0.59^{+0.10}_{-0.27}$&$0.720^{+0.041}_{-0.064}$&$0.729^{+0.064}_{-0.11}$\\
~~~~$\fave$\dotfill &Incident Flux (\fluxcgs)\dotfill &$1.347^{+0.085}_{-0.076}$&$0.697^{+0.040}_{-0.035}$&$0.295^{+0.018}_{-0.016}$\\

\label{bigtable}
\end{deluxetable*}




\end{document}